%% file: a478_xray.tex
\documentclass[longnamesfirst]{emulateapj}
\usepackage{apjfonts}

\shortcites{allen03,birzan04,boh02,chu01,chu02,dePlaa04,ettori02,fab03,
 fab00,fai00,fal05,fin04b,fin04a,forman04,garilli96,hansen05,jon02,
 kaastra04,lumb03,makishima01,mazzotta04b,mazzotta02,mcn01,moo98,nulsen05,
 omma04,osu03b,pat97,peterson03,piffaretti05,poi04,rizza00,rus04,rus04b,
 sand04,san03,schuecker04,sun03,tamura01,vikhlinin05,vik98b,zha04}

\defcitealias{sun03}{S03}

\slugcomment{Accepted by ApJ}  
\newcommand{\rmsub}[2]{\ensuremath{#1_{\mathrm{#2}}}} 
\newcommand{\srel}[2]{\mbox{\ensuremath{#1 - #2}}} 
\newcommand{\ps}{\ensuremath{\mathrm{s}^{-1}}}
\newcommand{\kmps}{\ensuremath{\mathrm{km}~\ps}}
\newcommand{\TX}{\rmsub{T}{X}}
\newcommand{\LT}{\srel{L}{\TX}}
\newcommand{\Mgas}{\rmsub{M}{gas}}
\newcommand{\MgasT}{\srel{\Mgas}{\TX}}
\newcommand{\Msol}{\ensuremath{\mathrm{M}_{\odot}}}
\newcommand{\omegal}{\rmsub{\Omega}{\Lambda}}
\newcommand{\omegam}{\rmsub{\Omega}{m}}
\newcommand{\rhocrit}{\rmsub{\rho}{crit}}
\newcommand{\rs}{\rmsub{r}{s}}   
\newcommand{\ASCA}{\textit{ASCA}}
\newcommand{\Chandra}{\textit{Chandra}}
\newcommand{\eg}{{\textrm e.g.}}
\newcommand{\ROSAT}{\textit{ROSAT}}
\newcommand{\xmm}{\textit{XMM}}
\newcommand{\XMM}{\textit{XMM-Newton}}
\newcommand{\chisq}{\ensuremath{\chi^2}}
\newcommand{\nH}{\rmsub{N}{H}}

\shorttitle{AGN Heating in A478}
\shortauthors{Sanderson, Finoguenov \& Mohr}

\begin{document}

\title{Possible AGN Shock Heating in the Cool Core Galaxy Cluster Abell~478}

\author{Alastair J.~R. Sanderson\altaffilmark{1}}
\email{ajrs@astro.uiuc.edu}

\author{Alexis Finoguenov\altaffilmark{2}}

\author{Joseph J. Mohr\altaffilmark{1,3}}

\affil{\scriptsize 1) Department of Astronomy, University of Illinois,
 Urbana, IL 61801}
\affil{\scriptsize 2) Max-Planck-Institut f\"ur extraterrestrische Physik,
 Giessenbachstra\ss e, 85748 Garching, Germany}
\affil{\scriptsize 3) Department of Physics, University of Illinois,
 Urbana, IL 61801}


\begin{abstract}
We present a detailed X-ray study of the intracluster medium (ICM) of
the nearby, cool-core galaxy cluster Abell~478 ($z=0.088$), based on
\Chandra\ and \XMM\ observations. Using a wavelet smoothing hardness
analysis, we derive detailed temperature maps of A478, revealing a
surprising amount of temperature structure for an apparently well relaxed
cluster. We find the broad band \Chandra\ spectral fits yield temperatures
which are significantly hotter than those obtained with \XMM, but the Fe
ionization temperature shows good agreement. We show that the temperature
discrepancy is slightly reduced when comparing spectra from regions
selected to enclose nearly isothermal gas. However, by simulating
multi-temperature spectra and fitting them with a single temperature model,
we find no significant difference between \Chandra\ and \XMM, indicating
that non-isothermality cannot fully explain the discrepancy.

We have discovered four hot spots located between 30--50 kpc from the
cluster center, where the gas temperature is roughly a factor of 2 higher
than in the surrounding material. We estimate the combined excess thermal
energy present in these hot spots to be ($3\pm1) \times 10^{59}$ erg. The
location of and amount of excess energy present in the hot spots are
suggestive of a common origin within the cluster core, which hosts an
active galactic nucleus. This cluster also possesses a pair of X-ray
cavities coincident with weak radio lobes, as reported in a previous
analysis, with an associated energy of less than 10\% of the thermal excess
in the hot spots. The presence of these hot spots could indicate
strong-shock heating of the intracluster medium from the central radio
source -- one of the first such detections in a cool core cluster. Using
the high resolution of \Chandra, we probe the mass distribution in the core
and find it to be characterized by a logarithmic slope of $-0.35\pm0.22$,
which is significantly flatter than an NFW cusp of -1 and consistent with
recent strong lensing results for a number of clusters.

\end{abstract}



\section{Introduction}
\label{sec:intro}
Clusters of galaxies are critical sites for investigating the interaction
between galaxies and their environment -- particularly their impact upon
the gaseous intracluster medium (ICM). There is substantial evidence to
demonstrate that the ICM has been subjected to non-gravitational heating
and/or cooling, so as to break the simple, self-similar scaling of cluster
properties with mass; \eg\ the \LT\ \citep[\eg][]{edg91,arn99,fai00},
\MgasT\ \citep{mohr99} and isophotal size-temperature \citep{mohr97}
relations. Furthermore, it is apparent that the ICM is systematically
under-dense and more extended in less massive halos
\citep[\eg][]{pon99,san03,osmond04,afshordi05}, and there is a 
increasing excess of entropy in the gas in cooler systems \citep{pon03}.

While these observations can be explained by the impact of feedback
associated with galaxy formation \citep[\eg][]{voit03}, the precise details
of the mechanisms which mediate this interaction remain unknown. Of
critical importance is the role of radiative cooling, which could fuel star
formation and gradually deplete the reservoir of gas in the cluster
core. Despite the inherently unstable nature of cooling by thermal
bremsstrahlung, gas in the dense, undisturbed cores of galaxy clusters
appears not to be cooling at the expected level
\citep[\eg][]{tamura01,peterson03,kaastra04}, leading to greatly reduced mass 
deposition rates \citep{makishima01,boh02}. A plausible explanation for
this behaviour is that thermal conduction acts to transfer significant
amounts of energy from the outer regions of the ICM
\citep{kim03,voigt04}. However, this process cannot operate at all
\citep{kho04} or effectively enough
\citep[\eg][]{wise04} to heat the core in all cases, especially at lower 
temperatures.

Since many ``cooling flow'' clusters are known to harbor central radio
sources \citep{burns90,eilek04}, it is likely that active galactic nuclei
(AGN) can deposit energy directly into the ICM, and thus offset at least
some of the cooling. Recently, \citet{croston05} have demonstrated that
heating from low power radio galaxies could account for the steepening of
the \LT\ relation in X-ray bright galaxy groups. Although clusters are much
more massive, the amount of energy output by powerful radio sources is
certainly sufficient to heat the gas by the required amount \citep{chu02};
what is lacking is a clear understanding of exactly how this energy is
coupled to the ICM.

A number of clusters have been shown to possess cavities, where expanding
radio lobes have displaced the ambient medium and left a depression in the
projected X-ray surface brightness \citep[\eg\ see the compilation of][and
references therein]{birzan04}. However, the coolest gas is generally found
beside these lobes \citep{fab00,mcn01} and their essentially subsonic
expansion is not obviously heating the ICM by a significant amount. In the
meantime, the role of AGN driven bubbles in modifying the ICM is opening up
as a very promising avenue of research
\citep[see][for a recent review]{gardini04b}.

Alternatively, energy may be transferred to the ICM by means of acoustic
waves generated by turbulence \citep{fujita04}, or viscous dissipation of
sound waves originating from a central radio source \citep{rus04,rus04b}.
\citet{fab03} have recently discovered ripples in the X-ray emission from 
the Perseus Cluster, which point to a continuous energy output from the
central radio source that can balance cooling within the innermost 50 kpc
of the core.

In this paper we examine in detail the properties of the ICM in a nearby
galaxy cluster, using X-ray data from two different telescopes. This
apparently well relaxed system has a large cool core and its central galaxy
hosts an active radio source, affording an excellent opportunity to study
its interaction with the ICM. We assume the following cosmological
parameters: $H_{0}=70$ km\,s$^{-1}$\,Mpc$^{-1}$, $\omegam=0.3$
$\omegal=0.7$. Correspondingly, at our adopted redshift for A478 of 0.088,
1\arcsec = 1.65 kpc. Throughout our spectral analysis we have used XSPEC
11.3.0, incorporating the default solar abundance table of
\citet{and89}. All errors are 1$\sigma$, unless otherwise stated.

\section{Data Reduction}

\subsection{\textit{Chandra} X-ray Data Reduction}
A 42.4 ks \Chandra\ observation of Abell~478 was made on January 29, 2001
using the Advanced CCD Imaging Spectrometer (ACIS) in FAINT mode. The data
reduction and analysis were performed with Ciao version 3.0.2 and CALDB
version 2.26. This release of the \Chandra\ CALDB calibration files was the
first to incorporate a full correction for the degradation in quantum
efficiency of the ACIS detectors. Therefore, no further steps were taken to
correct for its effects.

Three separate light curves were constructed for the S1, S3 and the
remaining CCDs combined, using the recommended energy and time binning
criteria\footnote{http://cxc.harvard.edu/contrib/maxim/bg.}. A small flare
was identified in the light curve from the S1 CCD and another was found in
the combined I2/I3/S2/S4 light curve: both were excluded, leaving 40.7 ks
of useful data. Only events with \ASCA\ grades 0,2,3,4,6 were used, and bad
columns and hot pixels were excluded. In addition, those events associated
with cosmic ray afterglows were identified and removed. A correction was
applied to the level 1 events file to allow for the ACIS time-dependent
gain variation, using the tool
``corr\_tgain''\footnote{http://asc.harvard.edu/cont-soft/software/corr\_tgain.1.0.html},
and a new level 2 events file was generated by reprocessing this modified
level 1 events dataset.

Since A478 fills much of the \Chandra\ field of view, we use the Markevitch
blank-sky datasets to estimate the background
contribution\footnote{http://cxc.harvard.edu/contrib/maxim/acisbg.}. To
allow for small variations in the particle background level between the
blank sky fields and the A478 observation, we rescaled the effective
exposure of the background datasets according to the ratio of count rates
in the particle-dominated 7--12 keV energy band for the S1 CCD, which is
furthest away from the cluster. To avoid the bias caused by the presence of
contaminating point sources in the A478 dataset, we identified and excluded
such features using the following iterative scheme.

Images were extracted in the 0.5--2.0 keV band for the S1 CCD in both the
main (A478) and blank sky datasets. A background image was created by
smoothing the blank sky field with a Gaussian of width 1 arcmin, to filter
out Poisson noise. The main image was then searched for sources with the
ciao task WAVDETECT, using this background. The source regions found were
then masked out of both the main and blank sky datasets, and the remaining
counts in the 7--12 keV band were summed. A rescale factor was then
determined as the ratio of the net count rate in the background dataset to
that in the main dataset. The effective exposure time of the blank sky
dataset was then multiplied by this rescale factor and the process
repeated, until no new source regions were found. A total of 6
contaminating source regions were identified and excluded from the S1 CCD
data in this way. We found that the blank sky data were 8\% higher than in
the A478 observation, consistent with the value of 10\% quoted by
\citet[][hereafter \citetalias{sun03}]{sun03} for the same observation.

\subsection{\textit{XMM} X-ray Data Reduction}
\XMM\ observed A478 in orbit 401 (observation ID 0109880101), using
the thin filter to block visible light. In the analysis that follows we
present the data obtained with the pn-CCD camera \citep{xmm_pn_short},
operated in extended full-frame mode, with an exposure time of 42 ks. The
initial stages of the data reduction were performed using XMMSAS
5.4.1. Screening of the data for contamination from flares was achieved by
examining a light curve of detector counts in the 10--15 keV band. We have
adopted the approach based on the analysis of the count rate histogram, as
described in \citet{zha04}, which is more sensitive to the background
conditions during the observations than using a fixed threshold to identify
good time intervals.

With these screened photon event files we produced a pn image of A478 in
the energy bands 0.5--2.0 and 2.0--7.5 keV. In the imaging analysis, we
included photons near the pn-CCD borders, near bad pixels and offset
columns, in order to reduce the width of the gaps between CCD chips. This
is a reasonable approach to adopt when dealing with qualitative images in
broad energy bands. For the spectral fitting, described in
\S\ref{sec:xray_spec}, we exclude all these border events, since a small
fraction of those photons have an incorrect energy, typically due to the
registration of a double event pattern as a single event.  After
subtracting the expected out-of-time events from the image, we corrected
the residual for vignetting and exposure using the latest calibrations
\citep{lumb03}, which have been incorporated within the XMMSAS 6.0 release.

Emission from A478 fills the entire \XMM\ field of view and so it was not
possible to determine the background level reliably from this dataset.
Therefore, an observation of the \Chandra\ Deep Field South was used to
provide an estimate of the background. We have confirmed that this choice
of dataset is well matched to the detector background in the 10--15 keV
band for our observation of A478. However, for the goal of the current
analysis the choice of the background is less important, since the measured
flux is strongly dominated by emission from the cluster.

\section{Chandra Spectral Analysis}
\label{sec:xray_spec}
A global spectrum of A478 was extracted from the emission filling the
entire S3 chip, excluding contaminating sources. The spectrum was grouped
to a minimum of 20 counts per bin, to enable us to use the
\chisq\ fit statistic. The CCD detector responses vary as a function of 
position within our spectral extraction region. We therefore generated
composite response files using the CIAO tasks MKWARF and MKRMF, by
averaging the contributions made from those parts of the detector with
different responses, weighting each component according to the distribution
of counts in the 0.5--2.0 keV band.

Using XSPEC 11.3.0, a single temperature MEKAL hot plasma model combined
with a WABS galactic absorption component, was fitted to the data between
0.7 and 7.0 keV, to maximize the signal-to-noise ratio (S/N). This yielded
a best-fit temperature of $6.56\pm0.06$ keV, with a metallicity of
$0.33\pm0.01$ solar, and an absorbing column of
$(2.91\pm0.01)\times10^{21}$ cm$^{-2}$.  The redshift was also left free to
vary, giving a best-fit value of $0.0879_{-0.0012}^{+0.0003}$, which agrees
well with the optical value of $0.0881\pm0.0009$ \citep{struble99}. We note
that \citetalias{sun03} found the following spectral parameters based on
their 0.7--8.0 keV analysis of data from the S2, S3 \& S4 CCDs combined
using this observation: $kT=7.18\pm0.11$ and
$\nH=(2.59\pm0.03)\times10^{21}$ cm$^{-2}$, with an abundance of
$0.37\pm0.02$ solar (all at 90\% confidence). The differences are not
surprising, since the gas temperature shows a marked decrease in the
cluster center and the metallicity and absorbing column both exhibit
central enhancements (see $\S$\ref{sec:ch_vs_xmm}) beyond the S3 chip, to
which we restrict our analysis. However, the significant discrepancy in
absorbing column is mainly due to the improvement in the \Chandra\
calibration since the \citeauthor{sun03} analysis -- a factor which we will
return to in Section~\ref{sec:deproject}.

\subsection{\textit{Chandra} vs \textit{XMM}}
\label{sec:ch_vs_xmm}
\begin{figure}
\hspace{5mm}
\includegraphics[width=7.5cm]{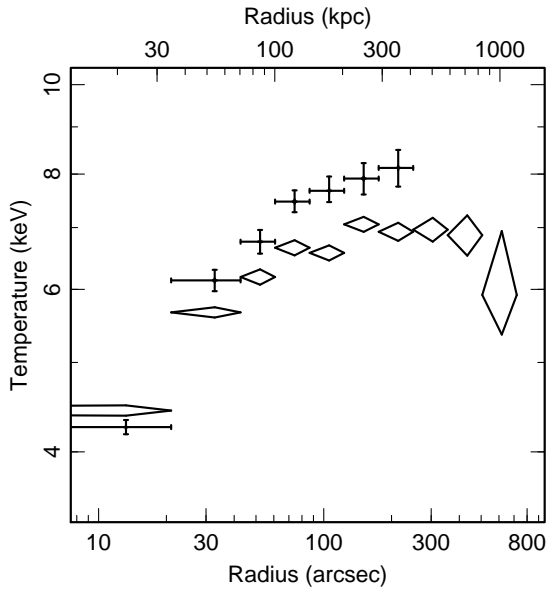}
\caption{ \label{fig:kT_2d}
Projected gas temperature as a function of radius. The diamonds are
the \XMM\ pn data and the barred crosses are the \Chandra\ data.}
\end{figure}

In view of our intention to exploit the complementary properties of 
\Chandra\ and \XMM, by combining spectral data from both (see 
$\S$\ref{sec:deproject}), we present here a detailed comparison of results
obtained from the two observatories. We are further motivated by the
apparent discrepancy between the temperatures inferred from \Chandra\ and
\XMM\ observations of A478, as reported by \citet{poi04}.  Given the
potential for spatial variation in cluster gas properties (as well as in
the absorbing column), we perform a direct comparison of spectral
properties in a series of contiguous annuli. Ten logarithmically-spaced
radial bins were defined out to a maximum radius of 12\arcmin\ (see
Table~\ref{tab:specfit}), centered on the peak of the emission as measured
using \Chandra\
(R.A. 04$^{\mathrm{h}}$13$^{\mathrm{m}}$25$^{\mathrm{s}}$.2,
decl. 10\degr27\arcmin53\arcsec). The innermost bin radius was increased to
21\arcsec, to reduce the effect of scattering between bins, due to the
comparatively broad point spread function (PSF) of \XMM. Equivalent
\Chandra\ spectra were extracted in the 7 innermost annuli fully covered by
the S3 chip.

\input{table1}  

\begin{figure}
\hspace{5mm}
\includegraphics[width=7.5cm]{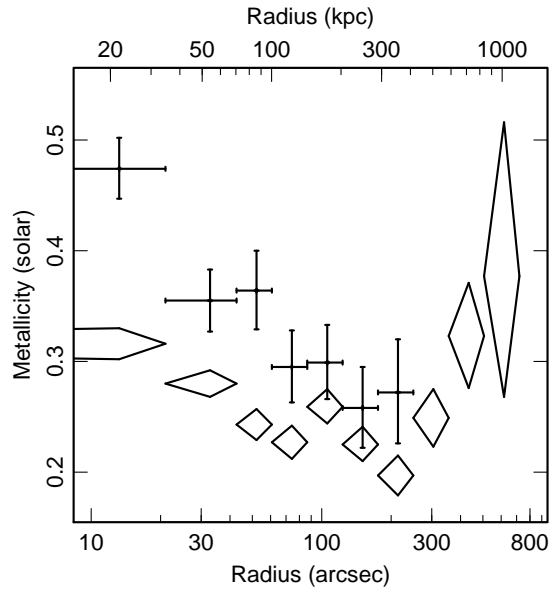}
\caption{ \label{fig:Z_2d}
Projected gas metallicity as a function of radius. The diamonds are
the \XMM\ pn data and the barred crosses are the \Chandra\ data. }
\end{figure}

\begin{figure}
\hspace{5mm}
\includegraphics[width=7.5cm]{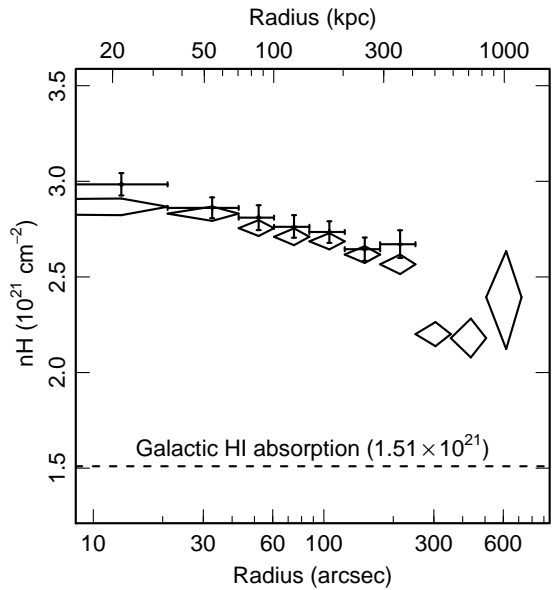}
\caption{ \label{fig:nH_2d}
Projected absorbing column as a function of radius. The diamonds are
the \XMM\ pn data and the barred crosses are the \Chandra\ data. The HI
inferred absorption value is indicated by the dotted line.}
\end{figure}

Absorbed MEKAL models were fitted to each spectrum, as before, and radial
profiles of temperature, metallicity and absorbing column are plotted
in Figs.~\ref{fig:kT_2d}--\ref{fig:nH_2d}, respectively, showing the
\Chandra\ data as barred crosses and the \XMM\ data as diamonds. The
temperature profile is typical of a `cool-core' cluster, showing a sharp
drop towards the center, within $\sim$1\arcmin. The region outside the core
is roughly isothermal, with some indication of a negative gradient further
out. It is also clear that the \Chandra\ points are systematically higher
in temperature compared to \XMM. This discrepancy persists in the measured
abundances, with \Chandra\ again giving somewhat higher values than \XMM\
in all annuli (Fig.~\ref{fig:Z_2d}). Nonetheless, both observations show an
increase in the gas metallicity in the cluster core, approximately
coincident with the observed decrease in temperature in this region. 
\citet{vikhlinin05} present a recent analysis of the A478 \Chandra\ data, 
which incorporates the latest calibration. Their projected temperature 
profile is similar to our own, although it extends beyond the S3 chip, 
reaching a slightly hotter peak of $\sim$9 keV at roughly 150 arcsec. 

Furthermore, the inferred absorption column also rises towards the center,
and there is close agreement between the \Chandra\ and \XMM\ results
(Fig.~\ref{fig:nH_2d}). Also plotted is the galactic absorption of
$1.51\times10^{21}$ cm$^{-2}$ based on HI observations \citep{dic90}
interpolated to the center of A478. It can be seen that the fitted points
lie well above this line, implying an observed absorption roughly twice as
high. However, this excess is not localized to the cluster core, and is
probably associated with absorption in our galaxy \citep{poi04}.  This
behavior is not unexpected, since HI measurements become an increasingly
unreliable predictor of the total galactic absorption for columns exceeding
$\sim$6$\times10^{20}$ cm$^{-2}$ due to contributions from molecular
hydrogen, for example \citep{lockman04}. An excess absorption was also
inferred from photometric observations by \citet{garilli96}, who noted that
the early-type galaxies in A478 are much redder than expected.

\begin{figure}
\hspace{5mm}
\includegraphics[width=7.5cm]{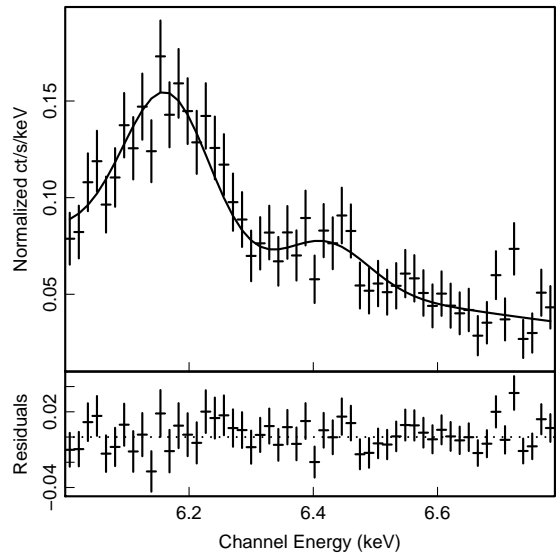}
\caption{ \label{fig:Fe_line}
 The 6.0--6.8 keV ACIS spectrum, best fitting MEKAL model and residuals
 from annuli 4--7 combined (see text for details).}
\end{figure}

In light of the disagreement in both temperature and metallicity between
\Chandra\ and \XMM, we performed additional tests to verify the impact of 
any calibration errors on our results. Specifically, we checked the
ionization temperature of the ICM, as measured from the prominent iron line
redshifted to $\sim$6.1 keV in the observer frame. We extracted a spectrum
from annuli 4--7 combined, thus excluding the inner cluster core, which
contains gas at a wide range of temperatures and metallicities, as already
seen. This spectrum was fitted with an absorbed MEKAL model as before, in
the ranges 0.7--7.0 keV and 6.0--6.8 keV (see
Fig.~\ref{fig:Fe_line}). Since the absorbing column was poorly constrained
when fitting the narrow (high) energy range, we fixed it at the value
obtained from the 0.7--7.0 keV fit. However, we also tried fixing
\nH\ at the galactic value deduced from HI observations
($1.51\times10^{21}$ cm$^{-2}$). This made no appreciable difference to the
results, which are summarized in the left half of
Table~\ref{tab:specrange}. The corresponding \chisq\ and degrees of freedom
are (711.07/427) and (39.49/51), for the broad and narrow range \Chandra\
fits, respectively. It is clear from the broad band fit that, even within
this apparently roughly isothermal region, the cluster emission is not well
described by a single temperature plasma. This hints at the presence of
significant spatial variation in the temperature of the ICM -- a
possibility which we investigate further in
\S\ref{sec:specmap_deproj}.

\input{table2}  

For direct comparison, the equivalent test was performed on the \XMM\ pn
spectrum from annuli 4--7; the results are summarized in the right hand
side of Table~\ref{tab:specrange}. As with \Chandra, it was necessary to
fix \nH\ for the narrow band fit at the optimum value obtained from the
0.7--7.0 keV fit. Once again we checked that the results do not change
noticeably when \nH\ was fixed at the HI value. However, we did find that
freezing the redshift at the optical value of 0.088 produced an
unacceptable fit (\chisq/dof = 423.20/154), indicative of a small
calibration offset, also found by \citet{poi04} and
\citet{dePlaa04}. This is due to the uncertainty in the current CTI 
correction affecting the pn detector in extended full frame mode only
(K.~Dennerl 2004, private communication), a correction for which has only
just been incorporated into the XMMSAS 6.1 release. Leaving the redshift
free to vary produced a best-fit value of $0.074\pm0.001$. To provide a
fairer comparison with \Chandra\ in a region heavily dominated by emission
line flux, we therefore applied a small modification to the response matrix
file to improve the fit, using the XSPEC ``gain'' command. This produced an
acceptable fit with a \chisq/dof of (158.65/154), compared to
(1337.99/1261) for the broad range fit, without the gain modification. The
2 fitted gain parameters were frozen at their best fit values in the
calculation of MEKAL parameter errors.

Table~\ref{tab:specrange} shows that the broad-band \Chandra\ spectrum
produces an anomalously high temperature, which disagrees with the
\Chandra\ ionization temperature measurement, as well as with both the
broad-band and ionization temperature fits from \XMM. This could indicate a
problem with the ACIS calibration as the source of the discrepancy between
\Chandra\ and \XMM\ reported by \citet{poi04} and seen in
Table~\ref{tab:specrange}, for this dataset. However, this behavior can be
explained by significant temperature variation within each annulus, giving
rise to different characteristic average temperatures when folded through
the \Chandra\ and \XMM\ responses and fit with a single phase plasma model
(see \S\ref{sec:specmap_deproj}). The implication is that better agreement
between \Chandra\ and \XMM\ ought to occur for spectra extracted from
approximately isothermal regions, which we find to be the case (see
$\S$\ref{sec:specmap_deproj}).

In the meantime, we have neglected to correct for this bias in what
follows, in view of the impracticality of restricting our \Chandra\
analysis to a small, high energy range, and given the somewhat better
agreement between \Chandra\ and \XMM\ seen in the deprojected temperature
profile (see $\S$\ref{sec:deproject}).

\subsection{The Effects of Multiphase Gas}
\label{sec:multiphase}
In order to shed some light on the discrepancy between \Chandra\ and \XMM\
described above, we have conducted a series of simulations of multiphase
spectra, which we have fitted with a single temperature spectrum. We have
used the spectral responses and background spectrum from annuli 4--7 (see
Table~\ref{tab:specrange}) for both telescopes and assumed a value of
galactic absorption of $2.7\times10^{21}$cm$^{-2}$ for all these
simulations.  Initially we generated 1000 realisations of a spectrum
comprising 4 temperature components (6,6.5,7.5 \& 8 keV, weighted in the
ratio 7:10:30:7), with an abundance of 0.3 solar. The results of fitting
this spectrum with a single absorbed MEKAL model recovered the following
mean temperatures and standard deviations: $7.11\pm0.10$ (\Chandra\
0.7--7.0 keV); $7.12\pm0.06$ (\xmm\ 0.7--7.0 keV); $7.16\pm0.45$ (\Chandra\
6.0--6.8 keV); $7.17\pm0.19$ (\xmm\ 6.0--6.8 keV). The excellent agreement
between these 4 cases demonstrates that moderately multiphase gas with the
same metallicity cannot account for the discrepancies seen in
Table~\ref{tab:specrange}, despite the differences in the spectral
responses for \Chandra\ and \XMM.

\input{table3}  

We then simulated a spectrum with 2 temperature components (5 and 12 keV)
and an abundance of 0.3 solar (case A), 0.2 \& 0.5 solar (case B), and 0.5 \&
0.2 solar (case C), respectively. The fitting results are summarised in
Table~\ref{tab:multiphase_sim}. It can be seen that there is consistently
good agreement between \Chandra\ and \XMM\ in all 3 cases. Furthermore, it
can also be seen that the 0.7--7.0 and 6.0--6.8 results are generally in
good agreement when both have the same metallicity. However, when the
hotter phase has higher metallicity, the narrow-band ionization balance fit
yields a significantly hotter temperature for both \Chandra\ and
\XMM. Conversely, when the cooler phase is more enriched, the 6.0--6.8 keV
fit yields a cooler temperature, albeit at rather lower significance.

As an aside we note that, in all cases, the metallicity obtained with a
single-temperature fit to the multiphase spectra overestimates the input
values, by $\sim$25\% in the simple case where the 2 phase have the same
abundance. Where the metallicities of the 2 phases differ, the abundance is
over-estimated by an even greater amount, for both the broad- and
narrow-band fits.

Based on these simulations, we therefore conclude that the discrepancy we
observe between \Chandra\ and \XMM\ cannot readily be explained by the
presence of multiple temperature components, even with differing
abundances.

\section{Deprojection Analysis}
\label{sec:deproject}
To determine the intracluster gas properties, we implemented a standard
``onion peeling'' scheme to deproject the X-ray data, under the assumption
of spherical symmetry. The spectral deprojection was performed using the
PROJCT model in XSPEC, using a single-temperature MEKAL plasma model
with a galactic absorption component (WABS). Only data within 0.7
and 7.0 keV were used, and each spectrum was grouped to a minimum of 20
counts per bin, including the background contribution.  The foreground
absorbing column density was left free to vary in each annulus, as was the
MEKAL temperature, abundance and normalization.

\begin{figure}
\hspace{5mm}
\includegraphics[width=7.5cm]{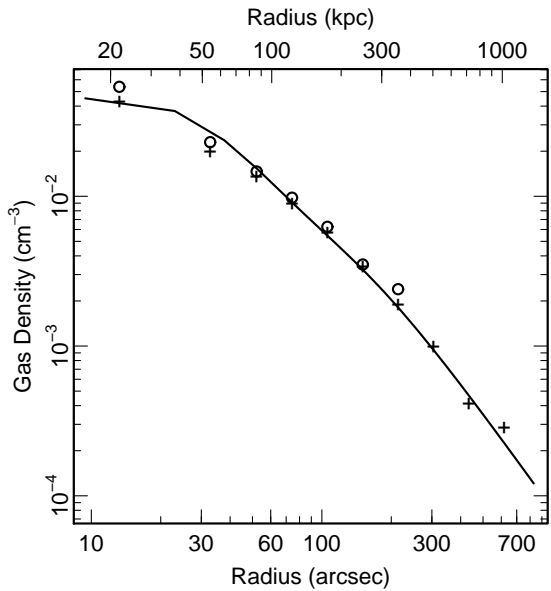}
\caption{ \label{fig:rho_3d}
 Gas electron number density as a function of radius. Errors on the density
 are too small to be plotted and the radial bin width error bars have been 
 omitted for clarity. The hollow circles represent the \Chandra\ points and
 the \XMM\ data are the ``+'' symbols. Also shown is the \ROSAT\ density
 profile from \citet{mohr99}, plotted as a solid line for clarity.}
\end{figure}

The XSPEC MEKAL normalization is defined as
\begin{equation}
K = \frac{10^{-14}}{4\pi (\rmsub{D}{A}(1+z))^2}\int{\rmsub{n}{e}\rmsub{n}{H} dV} ,
\label{eqn:mekal_norm}
\end{equation}
where \rmsub{D}{A} is the angular diameter distance, $z$ is the redshift
and $dV$ is the volume from which the deprojected emission originates. The
best-fit normalization for each spectrum was converted directly into a mean
gas density in the corresponding spherical shell, assuming a number density
ratio given by
\begin{equation}
\rmsub{n}{e}=\frac{\mu_{\mathrm{H}}}{\mu_{\mathrm{e}}} n_{\mathrm{H}} = 1.20 \, n_{\mathrm{H}} ,
\end{equation}
appropriate for a fully
ionized plasma with a metallicity of 0.3 times the solar value, using our
adopted abundance table from \citet{and89}.

Data from the outermost bin were discarded, since this volume incorporates
a contribution from emission beyond the last shell that implies a
non-trivial geometry for the purposes of determining the mean gas
density. This leaves a total of 6 \Chandra\ and 9 \XMM\ radial bins,
spanning the innermost 8.4\arcmin\ (835 kpc) of the ICM. Following
\citet{lew03} we assign an effective radius to each annulus, which is 
approximately equivalent to an emission-weighted mean, given by
\citep[see][]{mcl99}
\begin{equation}
r = \left[ 0.5\left( \rmsub{r}{out}^{3/2} + \rmsub{r}{in}^{3/2} \right) \right]^{2/3} .
\end{equation}
The resulting density profile is plotted in Fig.~\ref{fig:rho_3d}, and also
includes the data from the \ROSAT\ analysis of \citet{mohr99}, shown as the
solid line. There is generally good agreement between the 3 sets of data,
although the \Chandra\ points appear to be systematically higher than
\XMM. It is particularly encouraging to note the close agreement between
the \Chandra/\XMM\ values and the \ROSAT\ profile, since the latter was
derived from an analytical (double $\beta$-model with PSF corrections) fit,
as compared to the non-parametric ``onion peeling'' method used in this
analysis.

\begin{figure}
\hspace{5mm}
\includegraphics[width=7.5cm]{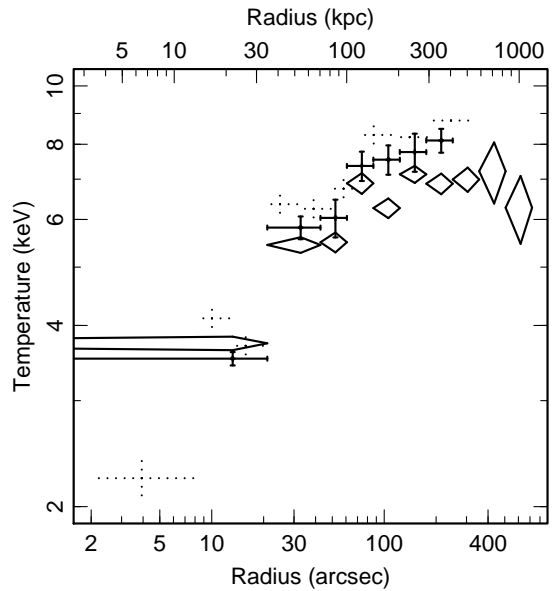}
\caption{ \label{fig:kT_3d}
 Deprojected temperature as a function of radius. The diamonds represent
 the \XMM\ pn data and the barred crosses are the \Chandra\ data. The
 dotted crosses are from the \citetalias{sun03} analysis of the same
 \Chandra\ observation, using older calibration data. Although some
 discrepancy between \Chandra\ and \XMM\ remains, it has been reduced with
 improvements to the CALDB and is less pronounced than in the projected
 $T(r)$ (Fig.~\ref{fig:kT_2d}).}
\end{figure}

The corresponding deprojected gas temperature profile is shown in
Fig.~\ref{fig:kT_3d} with the \Chandra\ and \XMM\ points plotted as before.
It can be seen that the \Chandra\ temperatures are hotter, except for the
innermost bin, where PSF scattering acts to smooth out the gradient in the
\XMM\ data. However, the agreement between the 2 observations is better than 
in the projected $kT(r)$, with an overlap in the $1\sigma$ error bounds for
all but 3 annuli. The deprojected temperature gradient is steeper than that
observed in the projected profile (Fig.~\ref{fig:kT_2d}), as expected
(since projection smooths out any such gradients), and reaches a minimum
central value of $\sim3.5$ keV. Also shown, for comparison, are the points
from \citetalias{sun03} using the same \Chandra\ data, plotted as dotted
crosses. These temperature values are systematically hotter than our
measurements. We attribute this discrepancy to changes in the \Chandra\
calibration between CALDB 2.15 and CALDB 2.26, particularly with respect to
the method for handling the effects of the ACIS contamination, which causes
an energy-dependent absorption that can affect temperature estimates. As
can be seen, the newer ACIS calibration produces temperatures which are
more consistent with those from \XMM.

\section{Spectral Mapping Analysis}

\begin{figure*}
\centering
\includegraphics[width=15cm]{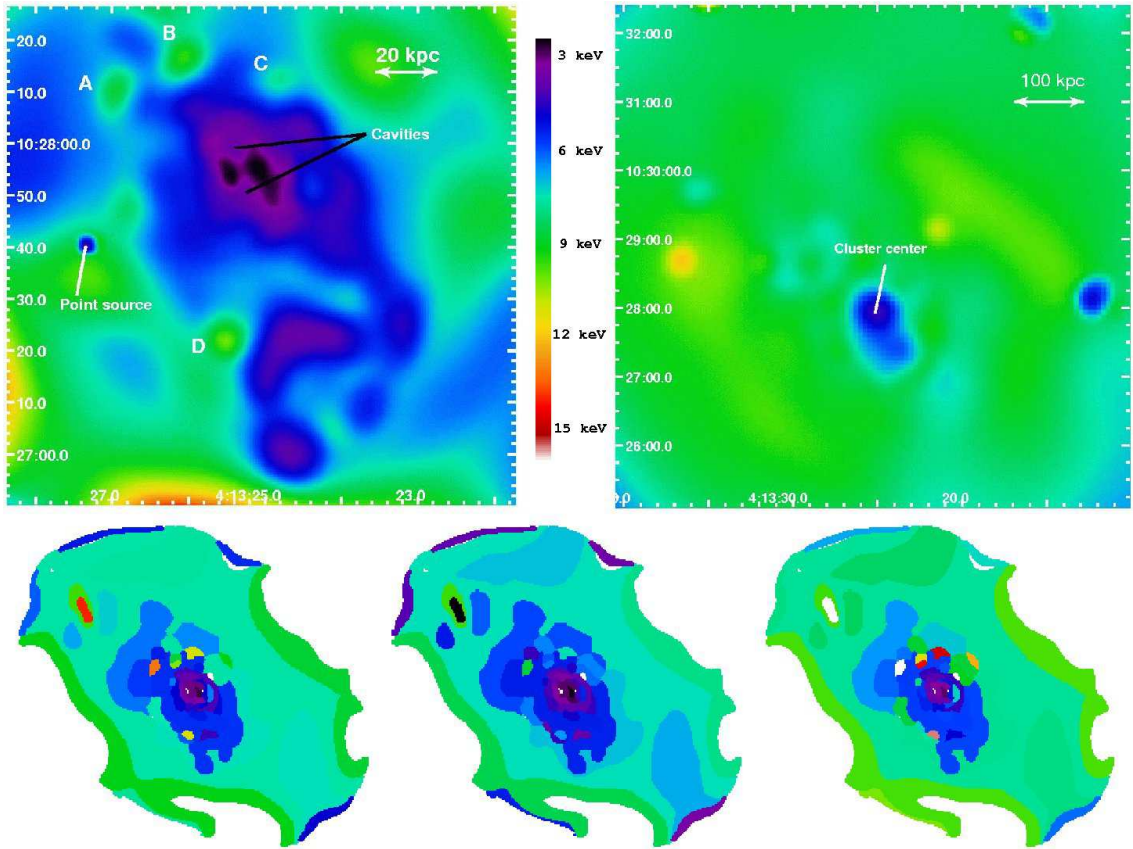}
\figcaption{Wavelet smoothed temperature maps of A478 from \Chandra\ (top 
 left) and \XMM\ (top right), derived from the hardness ratio method. The
 positions of 4 hot spot regions are indicated by the letters A-D, and the
 cavities discovered by \citetalias{sun03} are also labeled. The bottom
 half shows the \Chandra\ temperature map derived from the spectral fitting
 method; from left to right the panels show the best-fit values, 1$\sigma$
 low and upper limits on the temperature. The innermost green/yellow/orange
 regions are the hot spots.
\label{fig:kT_maps}}
\end{figure*}

\subsection{Hardness Method}
Although the surface brightness of A478 appears to exhibit a smooth
morphology typical of a relaxed cluster, it is difficult to gauge the true
state of the gas from studying images alone. Here we describe the process
of creating a temperature map from the \Chandra\ X-ray data, based on a
wavelet-smoothed hardness ratio image \citep{fin04a}. The key advantage of
this approach is that it allows fine scale temperature structure to be
resolved, without any prior knowledge of its morphology, since we are able
to work with high resolution images of the cluster.

We select two hardness bands, with the following energy ranges: 0.5--2.0
keV (soft band) and 2.0--7.0 keV (hard band). The hardness ratio (given by
the hard band flux divided by the soft band flux) depends strongly on the
gas temperature and the absorbing column along the line of sight, as well
as more weakly on the metallicity of the gas. Under the assumption that the
variation in absorbing column over the S3 chip is small (see the barred
points in Fig.~\ref{fig:nH_2d}), we can therefore use the hardness ratio to
estimate the temperature directly.

Taking the best-fit model to the global spectrum ($\S$\ref{sec:xray_spec}),
we fix the abundance and absorbing column and generate a series of 200 new
models with different temperatures in the range 0.5--15.0 keV. For each of
these spectra, we evaluate the ratio of the predicted model flux in the
hard to soft bands. We then fit a 6th order polynomial curve to the data,
to obtain an expression for the gas temperature as a function of
hardness. The residuals from this best-fit relation are less than 1\%
throughout the range of hardness values to which we apply it.

To create a hardness ratio map from X-ray data it is necessary to apply
some smoothing to suppress Poisson fluctuations. This ensures non-zero
pixel values in the denominator (soft band image) but can also reveal
subtle variations in hardness on small scales. We use a wavelet
decomposition approach, which is ideal for highlighting low surface
brightness diffuse emission, even in the presence of bright embedded point
sources, whilst retaining complex structural information \citep[see][for
details]{vik98b}. This approach also provides information on the
significance of structures identified on each scale and has recently been
used in the analysis of \XMM\ observations of several clusters
\citep{fin04a,fin04b,henry04}.

Initially we extract source and background images and corresponding
exposure maps in both the soft and hard bands for the whole S3 chip. The
corresponding background image is then subtracted from each hardness image,
and the resulting image divided by the appropriate exposure map. The
exposure maps are obtained by weighting the contributions to the effective
area of the detector by the best-fitting model to the global spectrum
fitted in $\S$\ref{sec:xray_spec}, within each of the soft and hard bands
separately. Pixels with an equivalent exposure less than 25\% of the
highest value within the S3 chip were masked out to improve the S/N of the
resulting hardness image.

A wavelet transform decomposition was then applied separately to the soft
and hard band images, over 6 different scales sizes, increasing in integer
powers of 2 from $2^2$ pixels to $2^7$ pixels, corresponding to angular
sizes of $\sim$2\arcsec\ and $\sim$1\arcmin, respectively. A 5$\sigma$
threshold was used to identify genuine sources, whose extent was defined to
enclose those neighboring pixels lying above a 2$\sigma$ threshold. Following
\citet{fin04a}, an additional smoothing was applied to each wavelet 
transform in turn, by convolving it with a Gaussian kernel of width equal
to that scale, to reduce discontinuity artifacts associated with separating
the image into different scales. The soft and hard band images were
reconstructed by summing all six smoothed wavelet transformed images
obtained at each of the scale sizes. The final hardness map was obtained as
the ratio of the hard to soft images. Each pixel was then converted into a
temperature value by using the polynomial relation derived above, with
unphysical hardness values (confined to a few pixels near the S3 chip
boundary) masked out of the resulting temperature map (i.e. values lying
outside the fitting range of the calibration curve).

\subsection{Spectral Fitting Method}
There are several key assumptions of the hardness-based approach to
generating a temperature map that could produce incorrect
results. Specifically, there may be subtle, small-scale variations in
absorbing column and gas metallicity that could manifest themselves as
spurious features in the inferred temperature distribution. Here we present
a more sophisticated method to determine spatial variations in gas
temperature, in order to assess the validity of the above technique.

With an estimated temperature map we are well positioned to identify
interesting features in the image for targeting with a conventional
spectral fitting approach. Using contour maps of the temperature and soft
band images, we defined a series of 53 separate regions, each containing a
minimum of 1000 counts in the raw image, originating from gas of similar
projected temperature and surface brightness, as implied by our wavelet
decomposition analysis.

We extract source and background spectra and response files in each of
these 53 regions, and fit each with an absorbed MEKAL model as
before. Images of these regions, showing the best-fit temperature and
$1\sigma$ bounds, are plotted in the lower half of
Fig.~\ref{fig:kT_maps}. The color table for the spectral fit temperature
map is identical to that used in the hardness temperature maps in the upper
half of the figure. The broad features in the hardness map are confirmed in
the spectral map: specifically, the temperatures of the cool core and
surrounding regions agree very well. Moreover, the spectral analysis
confirms the existence of 4 prominent hot spots located around the cluster
center, with temperatures as high as 12 keV -- almost double that of the
surrounding medium in all cases, albeit with large uncertainties. The
origin of these hot spots is unclear, but it seems unlikely that they can
all be attributed to point source contamination: visual inspection of the
raw image reveals no evidence for a surface brightness excess at these
locations.  We will return to the interpretation of these features in
$\S$\ref{sec:hotspot}.

\subsection{Spectral Mapping Deprojection}
\label{sec:specmap_deproj}
Using our spectral mapping results, it is possible to perform an
approximate deprojection, to provide a comparison with the
``onion-peeling'' method. To achieve this, we assume that all the emission
projected onto each region originates in a spherical shell bounded by the
inner- and outermost radii enclosing the region, measured from the surface
brightness centroid used previously. The volume used to calculate the gas
density from the MEKAL normalization (equation~\ref{eqn:mekal_norm}), is
formed from the intersection of this shell with the cylinder having the
cross section of that region. A more detailed description of this approach
is presented in \citet{henry04}.

In this ``deprojection'' scheme, no attempt is made to model out or
otherwise remove the contaminating emission from regions outside of the
volume element under consideration, which are projected onto the spectral
region. Therefore, we take the projected temperature and metallicity
information to apply to the 3-dimensional volume element. However, in
selecting these regions, we have ensured that that gas is essentially
isothermal as measured in projection, which minimizes the additional
smoothing effect of averaging over a range of temperatures. Correspondingly
the inferred temperatures and abundances are more likely to be
representative of the deprojected gas properties.

\begin{figure}
\hspace{-0.0cm}
\includegraphics[width=7.5cm]{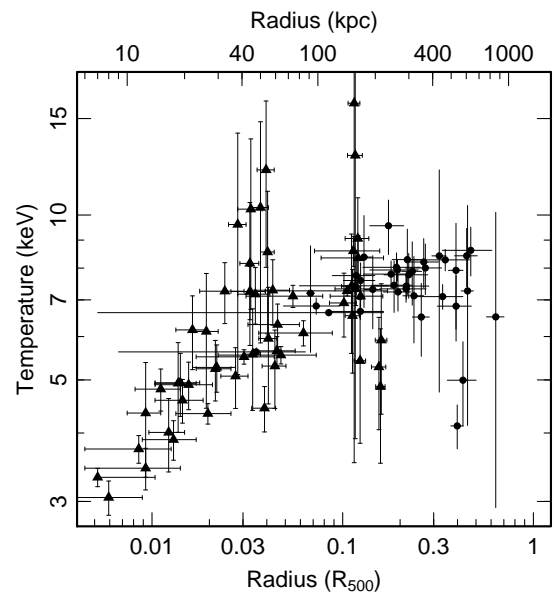}
\caption{ \label{fig:mapping_T(r)}
 Temperature profile from the spectral mapping analysis. No discrepancy
 between the \Chandra\ (diamonds) and \XMM\ (solid circles) points is
 evident in the mapping data.  }
\end{figure}

A plot of the temperature profile from the mapping analysis is presented in
Fig.~\ref{fig:mapping_T(r)}, with the radial axis expressed in units of
R$_{500}$ (1349 kpc), as determined from our mass model of the cluster (see
$\S$\ref{sec:M(r)}). It can be seen that there is a large degree of
temperature scatter, with the hot spots clearly visible, albeit with large
errors. The overlap between the \Chandra\ and \XMM\ spectral mapping data
is somewhat limited, due to the poorer spatial resolution of \XMM\ combined
with the impact of edge effects in the wavelet reconstruction limiting the
outer radius of data from the \Chandra\ S3 chip. However, it can be seen
that the discrepancy between \Chandra\ and \XMM\ found in the annular
spectral fits is not apparent in the spectral mapping data. Although this
is due in part to the larger uncertaintied in the data points, it may also
point to the source of the discrepancy being the varied temperature
structure in the ICM, which gives rise to a range of temperatures within a
given annulus. Since \XMM\ is more sensitive to cooler temperatures and
\Chandra\ is more sensitive to hotter ones, such a situation is likely to
lead to a bias in fitting a single temperature spectrum, as
observed. Moreover, the iron ionization fit is more sensitive to cooler gas
phases, which emit proportionately more line flux, which explains why the
\Chandra\ 6.0--6.8 keV result (Table~\ref{tab:specrange}) was in good
agreement with \XMM.

\begin{figure}
\hspace{5mm}
\includegraphics[width=7.5cm]{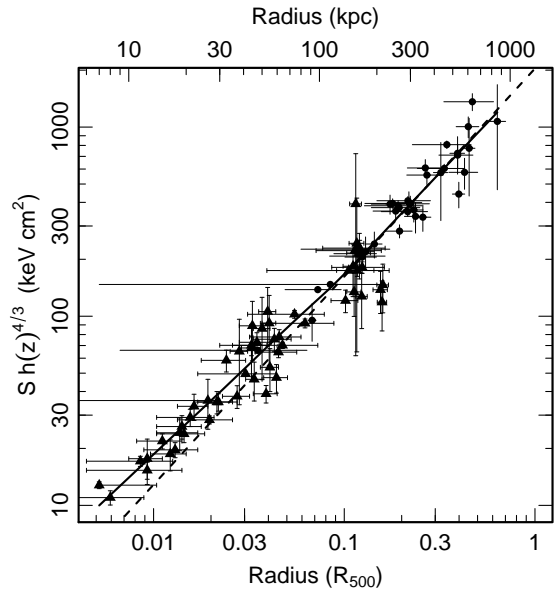}
\caption{ \label{fig:mapping_S(r)}
 Entropy profile from the spectral mapping deprojection. The point styles
 are the same as in Fig.~\ref{fig:mapping_T(r)}. The dashed line is an
 unweighted fit to the data, of the form $S\propto R^{1.1}$. For
 comparison, the solid line shows a locally weighted regression of the data
 in log-log space, using the method described in \S\ref{sec:hotspot}.}
\end{figure}

Fig.~\ref{fig:mapping_S(r)} shows the entropy profile from the spectral
mapping analysis. Also shown, as a dashed line, is a fit of the form $S
\propto r^{1.1}$ \cite[c.f.][]{toz00}, which has been shown to provide a 
good match to the outer regions of most reasonably relaxed clusters
\citep{pon03}. To highlight the trend in the data, the points have been
smoothed using a locally weighted regression in log-log space (solid
line). This technique smooths the data using a quadratic function which is
moved along the set of points to build up a curve, in an analogous fashion
to how a moving average is computed for a time series. The algorithm used
is implemented in the LOWESS function in the R Project statistical
environment package\footnote{http://www.r-project.org} (version 2.0.0)
\citep{Rcite}, and further details can be found there. The agreement in the
outer regions between this curve and the $S \propto r^{1.1}$ line is
striking, although the data deviate slightly above this simple relation in
the cool core (within $\sim$0.1 R$_{500}$ -- see
Fig.~\ref{fig:mapping_T(r)}). We defer further discussion of this issue to
\S\ref{sec:discuss_entropy}.

\subsection{Mass Distribution}
\label{sec:M(r)}
The deprojected gas $T(r)$ and $\rho(r)$ can be used to infer the 
gravitating mass profile, assuming hydrostatic equilibrium, given by
\begin{equation}
M_{\mathrm{grav}}\left(r\right)=-\frac{kT\left(r\right)r}{G\mu
 \rmsub{m}{p}}\left[\frac{\mathrm{d}\ln{\rho}}{\mathrm{d}\ln{r}}+\frac{\mathrm{d
}\ln{T}}{\mathrm{d}\ln{r}}\right],
\label{eqn:M(r)}
\end{equation}
\citep[e.g.][]{fabricant80}, where $\mu$ is the mean molecular weight of the gas and
\rmsub{m}{p} is the proton mass. In order to evaluate analytically the 
gradients in this equation, we fit a 3rd order polynomial to both the
$T(r)$ and $\rho(r)$ data in log space. The resulting mass profile is
evaluated at the radii of the input $T(r)$ and $\rho(r)$ values and is well
described by a profile of the form
\begin{equation}
\rho\left(r\right) = \frac{\rho_0}{x^n
   \left( 1+x \right)^{3-n}},
\label{eqn:gen_NFW}
\end{equation}
\citep[e.g.][]{zhao96}, where $\rho_0$ is the central density, $x=r/\rs$, 
and \rs\ is a characteristic scale radius. For $r \ll \rs$ the profile is
characterized by a cusp, with $\rho \propto
r^{-n}$. Equation~\ref{eqn:gen_NFW} is the generalized form of the
so-called NFW profile \citep{nav95}, which is obtained for $n=1$. A
reasonable proxy for the virial radius of the halo can be obtained from
$r_{200}$, the radius enclosing a mean overdensity of 200 with respect to
the critical density, $\rhocrit(z)$, of the Universe at the observed
cluster redshift, $z$, given by
\[
\rhocrit(z) =  E(z)^2\,\, \frac{3H_{0}^{2}}{8\pi G},
\]
where $G$ is the gravitational constant, $H_0$ is the Hubble constant at
$z=0$ and 
\[
 E(z) = (1+z) \sqrt{1+(z\,\,\omegam)+\frac{\omegal}{(1+z)^2}-\omegal},
\]
equal to 1.042 for our adopted redshift and cosmology. The concentration of
the NFW halo, $c$, is then obtained as $r_{200}$/\rs.

\begin{figure}
\hspace{5mm}
\includegraphics[width=7.5cm]{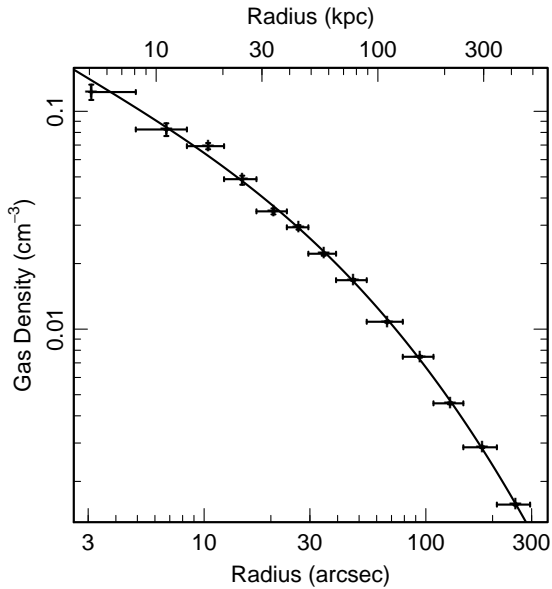}
\caption{ \label{fig:acis_rho_gas}
 High-resolution \Chandra\ gas density profile. The solid line is the 
 best-fit 3rd order polynomial relation in log-log space.
}
\end{figure}
\begin{figure}
\hspace{5mm}
\includegraphics[width=7.5cm]{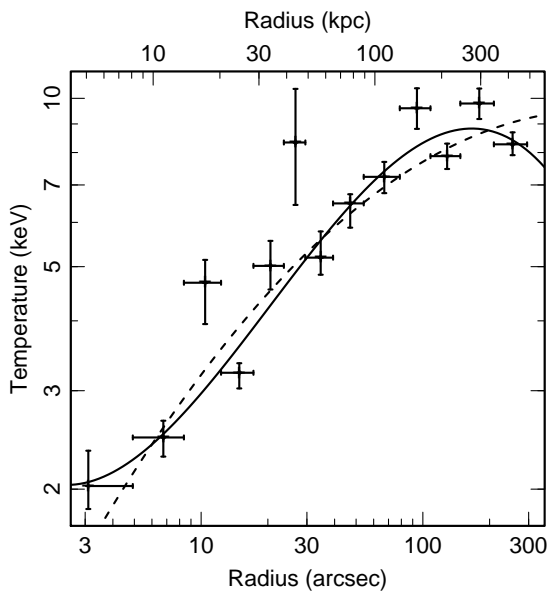}
\caption{ \label{fig:acis_kT_3d}
 High-resolution \Chandra\ gas temperature profile. The dashed and solid
 lines are the best-fit 2nd and 3rd order polynomial relations in log-log 
 space, respectively.
}
\end{figure}

\begin{figure}
\hspace{5mm}
\includegraphics[width=7.5cm]{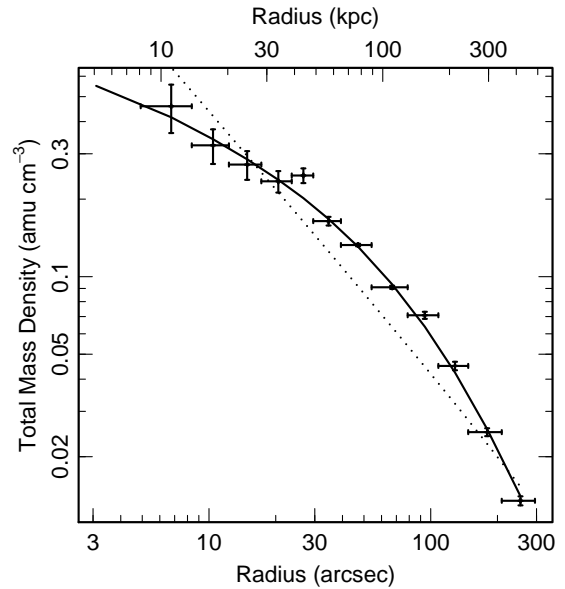}
\caption{ \label{fig:acis_M(r)}
 Mass profile from the \Chandra\ data. The solid line is the best fit using
 Equation~\ref{eqn:gen_NFW}, with an inner logarithmic slope of
 $-0.35\pm0.22$; for comparison, the dotted line shows the best-fit NFW
 profile (inner logarithmic slope of -1). The 1$\sigma$ errors were
 generated from a set of 2000 Monte Carlo simulations.}
\end{figure}

To provide better constraints on the mass distribution in the core of A478,
we perform a finer binned ``onion-peeling'' analysis using just the
\Chandra\ data, which has much higher spatial resolution compared to
\XMM. The resulting gas density and temperature profiles are plotted in
Figs.~\ref{fig:acis_rho_gas} \& \ref{fig:acis_kT_3d}, together with the
best-fit 3rd order polynomial curve (in log-log space) to the data. These
smoothed curves were used to evaluate the logarithmic gradients needed to
determine the mass profile using equation~\ref{eqn:M(r)}. Since we have
fitted these polynomials in log-log space, the logarithmic gradient is
simply the derivative of the function.

\begin{figure}
\hspace{5mm}
\includegraphics[width=7.5cm]{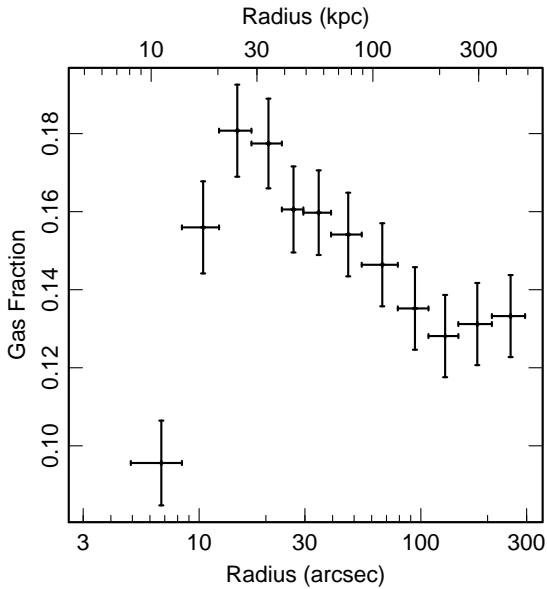}
\caption{ \label{fig:acis_fgas}
 \Chandra\ gas fraction profile, which shows an unusual decrease with
 radius within the cool core, beyond the central galaxy. The 1$\sigma$
 errors were generated from a set of 2000 Monte Carlo simulations.}
\end{figure}

The cumulative mass distribution was calculated at the radius of the input
$T(r)$ and $\rho(r)$ data, but using the values of the best-fit curve,
rather than the measured data points. We have excluded the innermost bin
since the gradient at this point is rather uncertain; we are more confident
about the outermost bin since the turn-over implied by the best-fit curve
is consistent with the temperature profile from \XMM, which extends beyond
this radius (see the diamonds in Fig.~\ref{fig:kT_3d}), and is also
confirmed by the \Chandra\ analysis of \citet{vikhlinin05}, which extends
beyond the S3 chip. The mass profile is plotted in
Fig.~\ref{fig:acis_M(r)}, together with the best-fit mass function obtained
with equation~\ref{eqn:gen_NFW}. The errors were determined from a series
of 2000 Monte Carlo realizations of the input $T(r)$ and $\rho(r)$ data; in
each case a best-fit mass model (equation~\ref{eqn:gen_NFW}) was fitted and
the errors on the model parameters were obtained from the standard
deviation of the set of points. Also plotted is the gas fraction profile
(Fig.~\ref{fig:acis_fgas}), which actually \emph{decreases} with radius out
to $\sim$200 kpc, beyond the confines of the central galaxy, somewhat
contrary to expectation \citep[e.g.][]{dav95,san03}.

We find a best fit index parameter of $n=0.35\pm0.22$, with a corresponding
scale radius and concentration parameter of
\rs\ = $317\pm82$ kpc and $c$ = $6.9\pm1.6$, giving a value of 
$R_{200}$ = $2190\pm125$ kpc and $R_{2500}$ = $620\pm70$
kpc. \citet{allen03} and \citet{schmidt04} also measured the \Chandra\ mass
profile for A478, but fitted it with an NFW profile (i.e. $n$ fixed at 1),
to give $R_{200}$ = $2464^{+190}_{-120}$ kpc (converted to $H_0$=70) and
$R_{200}$ = $2367$ kpc, respectively. \citet{poi04} also fitted an NFW
profile to the poorer-resolution \XMM\ data, to derive a value of $R_{200}$
= $2100\pm100$ kpc, in reasonable agreement with our findings. We estimate
a total mass within $R_{200}$ of $(1.3\pm0.27)\times10^{15}$ \Msol,
compared to $(1.84^{+0.48}_{-0.24})$ \Msol\ for \citeauthor{allen03}, and
$(1.1\times10^{15})$ \Msol\ for \citeauthor{poi04}. 

To investigate the robustness of our $M(r)$ inner logarithmic slope
measurement, we have also computed the mass profile for the case where a
2nd order polynomial is fitted to the temperature points (the dashed line
in Fig.~\ref{fig:acis_kT_3d}). In this case we recover a slightly steeper
index of $n=0.49\pm0.24$. However, we note that a 2nd order polynomial
indicates a rising $T(r)$ beyond the data points, whereas the 3rd order
polynomial curve exhibits a turn-over that is clearly preferred by our
\XMM\ data, as well as the \Chandra\ ACIS-I projected temperature profile of
\citet{vikhlinin05}. We return to discuss the mass distribution in
$\S$\ref{sec:discuss_M(r)}.

\section{Discussion}
\label{sec:discuss}

\subsection{Hot spots in the ICM}
\label{sec:hotspot}
While there is generally a surprising amount of temperature structure
present in the ICM in A478, of particular note are four ``hot spots'' seen
in the inner regions of the core (see Fig.~\ref{fig:kT_maps}). Three of
them are located to the North of the core; the other is to the South, and
they all lie within $\sim$30--60 kpc of the center. There are no known
cluster galaxies located at the position of these features, and they all
lie just outside the edge of the central cluster galaxy, PGC014685: its
$d_{25}$ diameter (the isophote at which the optical $B$ band surface
brightness reaches 25 mag/arcsec$^2$) is 57 kpc \citep{pat97}.

\begin{figure}
\hspace{5mm}
\includegraphics[width=7.5cm]{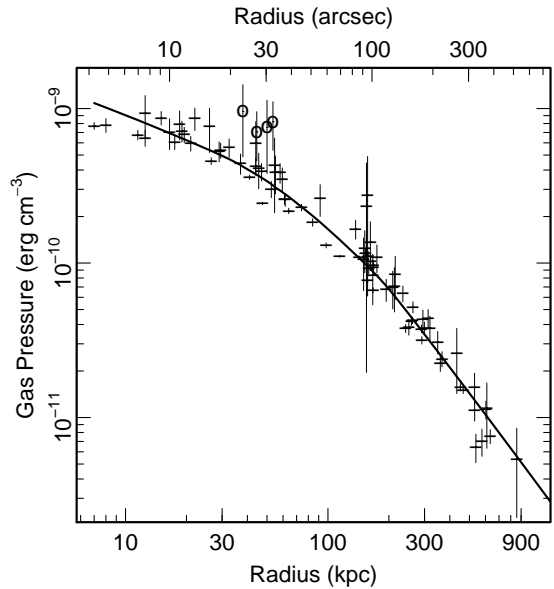}
\caption{ \label{fig:hotspot_P(r)}
 Pressure profile from the spectral mapping deprojection, with a smoothed
 curve shown as a solid line (see text for details). The hot spots are
 plotted as open circles. The X axis error bars have been omitted for
 clarity.}
\end{figure}

All four regions are clear outliers in both the temperature
(Fig.~\ref{fig:mapping_T(r)}) and pressure (Fig.~\ref{fig:hotspot_P(r)})
profiles, but occupy a relatively small volume. Correspondingly they
produce an essentially negligible distortion in the X-ray surface
brightness distribution of the core of A478. The entropy of the hot spots
also appears to be systematically higher than expected. Compared to the
smoothed regression relation plotted in Fig.~\ref{fig:mapping_S(r)}, there
is an excess entropy of between 20--50 keV cm$^2$ present in each of the
hot spots, corresponding to 30--50\% of the expected value. The combination
of excess pressure and entropy indicates that these are strong shocks, and
therefore unlikely to be the result of sound waves or mildly supersonic
motion. However, the volume enclosing the hot spots is too small for them
to produce a noticeable surface brightness contrast compared to the bright
emission from the cool core.

\input{table4}  

Some key properties of these hot spots are summarized in
Table~\ref{tab:hotspot}, including the estimated excess thermal energy
present in each one, based on the excess gas pressure and estimated
volume. This excess was measured relative to an estimate of the underlying
$P(r)$, obtained from a locally weighted regression of the data in log-log
space (the solid curve in Fig.~\ref{fig:hotspot_P(r)}), using the method
described in $\S$\ref{sec:specmap_deproj}. The excess energy in each hot
spot was calculated by multiplying its residual excess pressure above this
curve by the estimated volume for the region. Although the errors on the
individual estimates are quite large, the total excess thermal energy
contained in all 4 hot spots combined is ($3\pm1) \times 10^{59}$ ergs. We
note that this result depends only weakly on the volume, $V$, assumed for
the hot spots: the energy, $E \propto V\rho$, and $\rho\propto \sqrt{K/V}$
(Equation~\ref{eqn:mekal_norm}), where $K$ is the MEKAL model
normalization, which implies $E \propto \sqrt{KV}$.

The global morphology of A478 exhibits a regular structure, and the \XMM\
hardness temperature map (top right panel of Fig.~\ref{fig:kT_maps}) shows
no sign of significant disturbance in the outskirts of the ICM. This,
combined with the presence of a well established cool core, suggests that
A478 cannot have experienced a recent merger or significant disturbance,
which might account for the existence of the hot spots. It can be seen from
Table~\ref{tab:hotspot} that the distances from the cluster center and
energy excesses are similar for all the hot spots. Furthermore, they appear
to be located (see top left panel of Fig.~\ref{fig:kT_maps}) in rough
alignment with the (approximately) conical depressions discovered by
\citetalias{sun03}, corresponding to radio lobes emitted from the active
nucleus in the central galaxy. These characteristics point to the influence
of an event originating in the cluster center. Correspondingly, it is
reasonable to associate them with AGN activity.

On the basis that the hot spots share a common origin, we have extracted a
combined spectrum from all 4 regions and fitted it with a single absorbed
MEKAL model. In this way we can gain a better understanding of the degree
to which the gas has been heated. We obtain a best fit temperature of
$12.1^{+3.4}_{-2.6}$ keV; an abundance of $0.26^{+0.40}_{-0.26}$ solar and
aborption of $(2.4^{+0.3}_{-0.2}\times10^{21})$ cm$^{-2}$. However, by
fixing the temperature to the ambient value of 5.5 keV (e.g. see the
best-fit curve at $\sim50$ kpc in Fig.~\ref{fig:acis_kT_3d}) and refitting
with the other parameters left free to vary, we obtain a change in fit
statistic of $\Delta\chisq=16.7$, corresponding to 4.1$\sigma$
significance.

We stress that the interpretation of this result as evidence for the
existence of the hot spots hinges on the validity of associating these 4
separate regions with each other. It is certainly possible to achieve a
similar level of significance by randomly grouping hot regions arising from
purely statisitical fluctuations. However, we believe that the location of
the hot spots in the cluster core points to an association between them
that indicates they are not merely random peaks in the temperature
distribution. Specifically, they lie within a narrow annulus centered on
the cluster peak, and they are positioned within the range of angles that
enclose the roughly conical bi-polar cavities excavated by radio lobes from
the central AGN.

\subsection{AGN Heating?}
If the hot spots are indeed the result of AGN activity, we can address the
issue of how they may be connected to the cavities seen in the ICM. The
fact that the hot spots are located at least 3--4 times further away from
the cluster center implies a different formation mechanism to that which
excavated the cavities. In addition, \citetalias{sun03} estimate that the
minimum energy needed to create these cavities is $\sim$3$\times10^{58}$
ergs (\citet{birzan04} estimate $\sim$1.2$\times10^{58}$ ergs), which is a
tenth of the thermal energy contained in the hot spots. \citetalias{sun03}
further estimate a cavity age of $\sim$3$\times10^7$ yr, from the time
taken for the bubbles which formed them to rise buoyantly to their observed
location. Similarly, \citeauthor{birzan04} estimate a cavity age of between
1--3$\times 10^7$ yr, by considering 3 different timescales. For
comparison, the time taken to reach a distance of 50 kpc at the local sound
speed (920 \kmps) is $\sim$$5\times10^7$ yr, for a mean ambient gas
temperature of 5 keV (\rmsub{c}{s}=1480(\rmsub{T}{g}/10$^8$ K)$^{1/2}$
\kmps\ \citep{sar88}).  There is no evidence of a bow shock or X-ray
emission excess anywhere near these features, which argues against very
recent supersonic motion. It is more likely that the hot spots may have
been generated \textit{in situ}, perhaps in a similar manner to radio hot
spots located at the shock termination of radio jets, which have been
observed at similar distances from the cluster center \citep[\eg\ see the
recent sample of][]{hardcastle04}.

An estimate of the minimum time scale for heat conduction to erase the hot
spots can be obtained by considering the saturated heat flux, given by
\begin{equation}
q_{\mathrm{sat}} = 0.4\left( \frac{2 k T_{\mathrm{e}}}{\pi m_{\mathrm{e}}} \right)^{1/2}
 n_{\mathrm{e}} k T_{\mathrm{e}}
\end{equation}
\citep{cowie77}, where \rmsub{m}{e} is the electron mass. For hotspot A 
the gas density is $2\times10^{-4}$ cm$^3$ and its temperature is 12
keV. Assuming a spherical geometry for this region, with a volume of
$2.3\times10^{68}$ cm$^3$ (Table~\ref{tab:hotspot}), yields a cooling rate
of $10^{44}$ erg s$^{-1}$. Thus, it would take $\sim3\times10^7$ yr to lose
its excess energy, of 10$^{59}$ erg, via conduction. This is about a tenth
of the likely lifetime of the hot spots, which implies that heat conduction
must be suppressed by a factor of $\sim$10 compared to the Spitzer
rate. Such a decrease in conduction efficiency could easily be achieved by
magnetic fields, which have been shown to lead to suppression factors of 
$\sim$5 in a weakly collisional magnetized plasma \citep{nar01}.

The positions of the Northern set of 3 hot spots (Fig.~\ref{fig:kT_maps})
may offer a clue as to the mechanism responsible for their creation. Their
spacing is suggestive of a wide jet opening angle, of roughly 100
degrees. This is consistent with the slow jet heating scenario of
\citet{soker04}, which relies on such a wide angle to produce a poorly
collimated, massive and relatively slow outflow. Moreover, the emissivity
of the small bubbles in the ICM predicted from this model is too low for
them to be detected in X-ray images with existing telescopes, consistent
with the hot spots in A478. The \citeauthor{soker04} model is similar to
the effervescent heating model of \citet{begelman01}, which also postulates
the formation of many small bubbles, which heat the ICM by doing $PdV$ work
as they expand and rise.  A particularly attractive aspect of this scenario
is that the heating is smoothly regulated, since gas cooling acts to
steepen the ICM pressure profile, thus yielding more energy from the
bubbles' expansion so as to balance the cooling. However, this mechanism
for heating the ICM would need to be slow enough to avoid disrupting the
nearly power-law scaling of entropy with radius seen in
Fig.~\ref{fig:mapping_S(r)}.

While the interaction of AGN with the intracluster medium has been well
studied in recent simulations
\citep[\eg][]{rizza00,rey01,chu01,bru02b,basson03,rus04}, the formation of
shock heated blobs of the type seen in A478 does not appear to be a generic
feature. However, \citet{clarke97} present 3D magnetohydrodynamical
simulations of a supersonic jet, which predict the formation of a pair of
X-ray excesses, correspondingly to shocked ambient gas, at either side of
the jet orifice, which itself produces an X-ray cavity of the type seen in
A478. The \citeauthor{clarke97} simulations proceed for $\sim$10$^7$ yr,
which does not allow enough time to track the full evolution of these
excesses. However, it seems probable that the shock heated gas in these
blobs would expand outwards and heat the ICM, which could be an important
means of transferring AGN mechanical luminosity into gas thermal energy.

Similarly, the more recent simulations of \citet{omma04} appear to produce
three distinct density enhancements surrounding the head of an AGN outflow,
appearing after $\sim10^8$ yr. However, the jet responsible for these
features is only weakly relativistic, so may not be capable of heating
these blobs at the level observed in A478. Observationally, perhaps the
closest analogue to the hot spots reported here is the heated bubble
observed in MKW~3s by \citet{mazzotta02}. This feature lies at roughly 90
kpc from the cluster core and is approximately 25 kpc in radius in the
plane of the sky. More recently, \citet{mazzotta04b} present new, lower
frequency radio observations for this cluster, which show that the radio
emission is fully enclosed by the shocked gas. A similar radio/X-ray
interaction is observed in Hydra A, where expanding radio lobes, observed
at 330 MHz, appear to be driving a shock front \citep{nulsen05}.  It is
possible that equivalent lower frequency observations of A478 may reveal
larger radio lobes than those seen in the 1.4 GHz VLA map
\citep[shown in][]{sun03}, which might show some evidence of an interaction 
with the hotspot regions.

However, unlike A478, the MKW~3S bubble is clearly visible as a depression
in the surface brightness and the Hydra A shock front is revealed as a
discontinuity in the X-ray image. Similarly, there are spiral arm-like
features observed by \Chandra\ in the elliptical galaxy NGC~4636, produced
by weak shocks driven by off center outbursts \citep{jon02}. However, these
are ostensibly density enhancements, which produce a clear signature in the
surface brightness -- as already noted, the hot spots seen in A478 are
effectively invisible in X-ray images.

There is further evidence of possible AGN activity disturbing the gas in
the form of a tentative ``ripple'' in the surface brightness profile of the
core (Fig.~\ref{fig:unsharp_im}). We have generated an unsharp mask image,
using the method of \citet{rus04b}, who have simulated viscous dissipation
in the ICM caused by AGN activity. A 0.5--7.0 keV raw image was smoothed
with a Gaussian kernel of FWHM 6 kpc ($\sigma$=3.1 pixels) and subtracted
from the unsmoothed image; the resulting image has been smoothed with a
Gaussian kernel of $\sigma=1.5$ pixels to highlight the features
present. The depressions caused by the radio lobes are clearly seen,
producing an ``H'' shape which resembles that seen in the
\citet{rus04b} simulations, and there is also a suggestion of a ``ripple'' 
to the North of the center, lying between 16 and 26 kpc of the AGN. This
ripple has a low contrast, but is clearly longer and more coherent that the
typical noise features in the image. It is reminiscent of the ripples seen
in the ICM of the Perseus \citep{fab03} and Virgo (at 14 and 17 kpc)
\citep{forman04} clusters, which are attributed to shock fronts emanating
from the AGN in the central cluster galaxy. If this ripple is genuine, the
implied travel time for the feature is roughly 2$\times 10^7$ yr (given the
assumed local sound speed of 920 \kmps\ used above), which is comparable to
the age of the cavities.  We note that the hot spots in A478 are not
visible in the unsharp mask image and that they lie at least twice as far
out as the putative ripple.

\subsection{Entropy Profile}
\label{sec:discuss_entropy}
The radial variation in gas entropy in A478 (Fig.~\ref{fig:mapping_S(r)})
is unusual, in that it appears to follow a broken power law relation, with
an outer logarithmic slope of $\sim$1.1 and a slightly shallower inner
slope of $\sim$0.95 within $\sim$0.1 R$_{500}$, but with no indication of a
flattening in the core. Non cool-core clusters generally exhibit a
flattening of the entropy profile towards the center
\citep[\eg][]{ras04,pratt05}, while the effect of cooling is to lower the
entropy in exactly this region. It therefore appears to be a conspiracy
that causes these two effects to counterbalance each other so as to
preserve the scaling evident in the outskirts, where cooling is
insignificant. A recent \XMM\ study of 13 nearby cool core clusters has
found that their entropy profiles follow a power-law with best-fit
logarithmic slope of 0.95, out to roughly half the virial radius 
\citep{piffaretti05}. However, other cool core clusters show evidence of 
flattening in their core entropy profile
\citep[\eg][]{pratt03,osu03b,wise04}. Moreover, the slope of
$\sim$1.1 is predicted to arise from shock heating due to accretion of gas
\citep{toz00,toz01}, which would not be expected to persist in
cooling-dominated cluster cores.

The shape of A478's entropy profile resembles that of the fossil group
NGC~6482 ($kT\sim0.7$ keV), which was the subject of a recent \Chandra\
analysis \citep{kho04}. That observation is restricted to the inner 10\% of
$R_{200}$ of the halo, but is is clear that the entropy data are consistent
with a power law form, with a logarithmic slope slightly shallower than $S
\propto r^{1.1}$. Since NGC~6482 is a fossil group, it is expected to be a
very old system. However, it shows no evidence for strong cooling and,
furthermore, has a gas temperature profile which decreases monotonically
with radius, thus ruling out thermal conduction as a means of heating the
core. Since NGC~6482 shows some evidence of having a mildly active nucleus,
it is possible that this system too has been subjected to a phase of AGN
heating which has only recently abated, where significant gas cooling is
presently being reestablished.

\begin{figure}
\centering
\includegraphics[width=7.5cm]{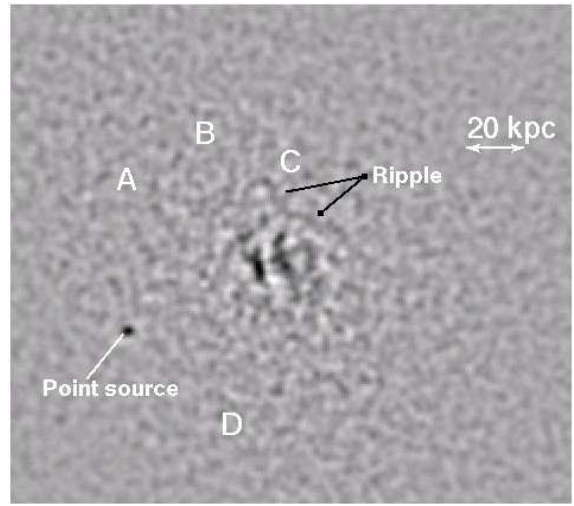}
\figcaption{An unsharp masked 0.5--7.0 keV \Chandra\ image of the core of 
 A478. Each pixel is 0.492\arcsec\ (0.8 kpc) across and the image has been
 smoothed with a Gaussian of $\sigma=1.5$ pixels. The edges around the
 radio lobe cavities are clearly visible as an ``H'' shape, and there is
 evidence of a possible curved ``ripple'' to the North of the cavities. The
 positions of the other features are labeled exactly as positioned in
 Fig.~\ref{fig:kT_maps}.
\label{fig:unsharp_im}}
\end{figure}

Given the apparently significant impact of the AGN on the gas in the core
of A478, it is worth noting that excessively large heat input would trigger
convective motions that would stir up the gas, flattening the entropy
profile and erasing abundance gradients in the core
\citep{chu01,bru02,chu02}. As there is some enhancement in the metallicity 
towards the center the ICM (Fig.~\ref{fig:Z_2d}), it is unlikely that a
significant disturbance could have occurred recently. This may be because
the AGN heating is widely distributed (e.g. effervescent) and thus not very
vigorous. Alternatively, it could indicate that the most unstable
(i.e. lowest entropy) gas is also the most metal rich, and that as it cools
and moves inwards it produces a central abundance peak.

The cooling time of gas in the center of A478 is only $\sim$9$\times 10^7$
yr --- roughly 6 times shorter than the cooling time of the hot spots (see
Table~\ref{tab:hotspot}). \citetalias{sun03} point out that the observed
radio lobes are much smaller than the X-ray cavities, which suggests a
fading of the radio source on a timescale of $\sim$10$^{7-8}$
yr. Furthermore, the ratio of mechanical power needed to excavate the
cavities to the current radio power of the AGN is over 1000
\citep{birzan04} -- by far the largest ratio amongst the 10 clusters (from
their sample of 18 objects) for which they have such data. The inclusion of
the hot spot contribution to the energy budget raises this value even
higher. Therefore, it is possible that we are witnessing the rapid
reestablishment of a cluster cooling flow, following a period of AGN
heating of the gas, before the cavities have been completely refilled.  On
the other hand, an AGN can vary in radio power while its jet power remains
constant \citep{eilek04,birzan04}, and thus some mechanical heating of the
ICM may be ongoing, despite the fading radio luminosity, as possibly
indicated by the putative ``ripple'' seen in Fig.~\ref{fig:unsharp_im}.

If similar hot spot features are present in other clusters with radio lobe
cavities, the timescale for which these features persist as over-dense
blobs is short enough to make them unlikely to be detected in X-ray images.
Furthermore, for much of this time they would reside in close proximity to
the cluster center, where emission from the cluster core would be
substantial, greatly reducing their surface brightness contrast. The
alternative method of detecting them -- locating the hot spots from their
temperature signature -- is also challenging, since it requires high
resolution hardness mapping and the best possible data quality (the A478
\Chandra\ observation analysed in this work comprises $\sim$400,000 net 
source counts from the S3 chip).

\subsection{Mass Profile}
\label{sec:discuss_M(r)}
The soft core in the mass profile (Fig.~\ref{fig:acis_M(r)}) of A478 is
unusual, and apparently at odds with numerical simulations of clusters,
which indicate inner logarithmic slopes of -1 \citep{nav95} or even as
steep as -1.5 \citep{moo98}. However, such a flat slope is permitted by
purely analytical considerations of the dark matter, based on fundamental
statistical mechanics: \citet{hansen05} have recently demonstrated that
this approach permits $0 < n < 10/3$, for the index parameter in
equation~\ref{eqn:gen_NFW}. Furthermore, a number of clusters have been
found with similarly flatter cusps in their mass profiles. \citet{tyson98}
measured a slope of $0.57\pm0.02$, and \citet{sand02} observed a value of
0.35 for two different clusters, based on strong gravitational lensing
analyses. More recently, \citet{sand04} have studied lensing arcs in 6
clusters and report a mean slope of $0.52\pm0.05$. It is interesting to
note that their sample comprises only clusters with a dominant brightest
cluster galaxy (BCG), as is the case for A478. Also, the cluster Abell~1795
has been shown to have an inner mass profile logarithmic slope of
$0.59^{+0.12}_{-0.17}$ (90\% confidence), inferred from a \Chandra\
analysis \citep{ettori02}.

It must be remembered that, unlike lensing measurements, our X-ray analysis
hinges on the assumption of hydrostatic equilibrium, which could easily
break down in the core, given the impact of the central AGN. For example,
the presence of significant non-thermal pressure support could mimic a less
concentrated mass distribution. The simulations of \citet{fal05} indicate
that random bulk motions of gas account for 10\% of the total pressure
support in clusters. Also, a recent \XMM\ analysis of the spectrum of
pressure fluctuations in the Coma Cluster found a lower limit of $\sim$10\%
for the contribution to the ICM pressure support arising from 
turbulence. It is possible that bulk or turbulent motion of gas associated
with AGN activity could be significant in the core of A478. Presumably such
a mechanism would give rise to a radially diminishing pressure
contribution, which could act to flatten the inferred mass distribution in
the manner observed. In such a situation, the inferred gas fraction would
be increasingly overestimated at smaller radii, which may explain the
radially decreasing trend observed within $\sim$200 kpc in
Fig.~\ref{fig:acis_fgas}.

\section{Conclusions}
We have studied the detailed thermodynamic properties of the intracluster
medium in the relaxed, cool-core cluster Abell~478, using \Chandra\ and
\XMM\ X-ray data. Our main findings can be summarized as follows:

\begin{enumerate}
\item We find that our \Chandra\ X-ray temperature measurements are 
 systematically hotter than those from \XMM, as also reported by
 \citet{poi04}. By selecting approximately isothermal regions, we find
 slightly better agreement between the two. However, by simulating
 multiphase spectra and fitting them with a single temperature model, we
 find no evidence for significant disagreement between \Chandra\ and
 \XMM. We therefore conclude that the observed discrepancy cannot be fully
 attributed to non-isothermality in fitting single-temperature models.

\item The entropy profile appears to agree well with the empirical modified 
 entropy scaling of \citet{pon03}. Moreover, the power law trend continues
 to the innermost radius measured ($<$10 kpc), with only a slightly 
 shallower slope. There is no evidence of any core in the entropy profile.

\item Under the assumption of hydrostatic equilibrium, we infer a mass 
 profile that exhibits a soft core, characterized by a logarithmic slope of
 $-0.35\pm0.22$. This is significantly flatter than an NFW profile (slope =
 -1), but is consistent with the recent gravitational lensing results of
 \citet{sand04} for clusters containing a dominant central galaxy.

\item We have discovered four hot spots in the ICM located well within the
 cool core, where the gas is roughly twice as hot as its surroundings. The
 combined excess energy associated with these regions is ($3\pm1) \times
 10^{59}$ erg, which is $\sim$10 times the energy needed to excavate 
 the cavities produced by radio lobes from the central AGN. The properties
 of these hot spots suggest they may the result of strong shock heating 
 from a jet/outflow originating in the AGN.
\end{enumerate}

\acknowledgments 
We are grateful to M. Arnaud and E. Pointecouteau for useful discussion and
providing the results of their \XMM\ analysis prior to publication. AS
thanks Yen-Ting Lin for useful discussions, Ming Sun for providing his
published Chandra profile data in electronic form, Ben Maughan for spotting
a mistake in Equation~\ref{eqn:mekal_norm} and the referee for helpful
comments; AF thanks The University of Illinois for hospitality during his
visit. This work is supported by NASA Long Term Space Astrophysics award
NAG5-11415.  This work has made use of the NASA/IPAC Extragalactic Database
(NED) and the HyperLeda galaxy database.


\bibliographystyle{apj}
\bibliography{$AJRS_LATEX/ajrs_bibtex} 

\end{document}

%% file: table1.tex
%

\begin{deluxetable*}{cccccccccc}
\tablecolumns{10}
\tablewidth{0pc}
\tablecaption{Spectral fitting results (0.7--7.0 keV)}
\tablehead{
\colhead{}    &  \multicolumn{3}{c}{\Chandra} &   \colhead{}   &
\multicolumn{3}{c}{\XMM} \\
\cline{2-5} \cline{7-10} \\
\colhead{\rmsub{R}{out}$^{a}$ (\arcsec)} & \colhead{$kT$ (keV)}   & \colhead{$Z$ (solar)}    & \colhead{$nH$ ($10^{21}$\,cm$^{-2}$)} &  \colhead{$\chi^2$/dof} &
\colhead{}    & \colhead{$kT$ (keV)}   & \colhead{$Z$ (solar)}    & \colhead{$nH$ ($10^{21}$\,cm$^{-2}$)} & \colhead{$\chi^2$/dof} }
\startdata
21	& $4.25\pm0.08$ & $0.47\pm0.03$ & $2.98\pm0.06$ & 459.8 / 346 && $4.43\pm0.06$ & $0.32\pm0.01$ & $2.87\pm0.04$ & 1367.8 / 1006 \\
43	& $6.14\pm0.16$ & $0.35\pm0.03$ & $2.86\pm0.05$ & 434.7 / 376 && $5.67\pm0.07$ & $0.28\pm0.01$ & $2.83\pm0.04$ & 1456.2 / 1150 \\
61	& $6.76\pm0.20$ & $0.36\pm0.04$ & $2.81\pm0.06$ & 415.3 / 357 && $6.19\pm0.12$ & $0.24\pm0.01$ & $2.76\pm0.04$ & 1312.2 / 1083 \\
87	& $7.47\pm0.20$ & $0.29\pm0.03$ & $2.76\pm0.06$ & 437.6 / 374 && $6.66\pm0.13$ & $0.23\pm0.01$ & $2.71\pm0.04$ & 1197.7 / 1095 \\
123	& $7.67\pm0.28$ & $0.30\pm0.03$ & $2.74\pm0.06$ & 455.3 / 379 && $6.57\pm0.12$ & $0.26\pm0.02$ & $2.69\pm0.04$ & 1325.2 / 1096 \\
175	& $7.91\pm0.31$ & $0.26\pm0.04$ & $2.64\pm0.06$ & 443.5 / 385 && $7.06\pm0.13$ & $0.23\pm0.02$ & $2.62\pm0.04$ & 1189.4 / 1116 \\
250	& $8.12\pm0.37$ & $0.27\pm0.05$ & $2.67\pm0.07$ & 454.1 / 379 && $6.93\pm0.16$ & $0.20\pm0.02$ & $2.57\pm0.05$ & 1039.3 / 1058 \\
355	& \nodata & \nodata & \nodata & \nodata && $6.96\pm0.21$ & $0.25\pm0.03$ & $2.20\pm0.06$ & 1025.5 / 1002 \\
506	& \nodata & \nodata & \nodata & \nodata && $6.87\pm0.35$ & $0.32\pm0.05$ & $2.18\pm0.11$ & 1095.8 / ~~975 \\
720	& \nodata & \nodata & \nodata & \nodata && $5.92\pm0.57$ & $0.38\pm0.11$ & $2.40\pm0.25$ & 1035.0 / 1030 \\
\enddata
\label{tab:specfit}
\tablecomments{$^{a}$Outer radius of annulus. The redshift was fixed at 0.088 for all spectra. All errors are $1\sigma$.}
\end{deluxetable*}
 

%% file: table2.tex
%

\begin{deluxetable*}{lccccccccc}
\tablecolumns{10}
\tablewidth{0pc}
\tablecaption{Spectral fitting results for annuli 4--7 combined}
\tablehead{
\colhead{}    &  \multicolumn{3}{c}{\Chandra} &   \colhead{}   &
\multicolumn{4}{c}{\XMM} \\
\cline{2-5} \cline{7-10} \\
\colhead{Fit Range (keV)} & \colhead{$kT$ (keV)}   & \colhead{$Z$ (solar)}    & \colhead{$nH$ ($10^{21}$\,cm$^{-2}$)} & \colhead{$\chi^2$/dof} &
\colhead{}    & \colhead{$kT$ (keV)}   & \colhead{$Z$ (solar)}    & \colhead{$nH$ ($10^{21}$\,cm$^{-2}$)} & \colhead{$\chi^2$/dof}}
\startdata
0.7--7.0 & $7.73\pm0.15$ & $0.28\pm0.02$ & $2.72\pm0.03$ & 711.1 / 427 && $6.81\pm0.06$ & $0.28\pm0.01$ & $2.66\pm0.02$ & 1338.0 / 1256 \\
6.0--6.8 & $6.49\pm0.61$ & $0.22\pm0.05$ & \hbox{}~~$2.72$ (frozen) & 39.5 / 51 && $6.69\pm0.26$ & $0.24\pm0.02$ & \hbox{}~~$2.66$ (frozen) & 158.5 / 152\\
\vspace{-0.6em}
\enddata
\label{tab:specrange}
\tablecomments{The redshift was fixed at 0.088 for \Chandra, but left free to vary for the \XMM, 
 broad-band fit; a simple linear gain offset was fitted to the narrow-band \XMM\
 spectrum to give the correct redshift (see text for details). All errors
 are 1$\sigma$.}
\end{deluxetable*}
 

%% file: table3.tex
%

\begin{deluxetable*}{ccccccccccc}
\tablecolumns{11}
\tablewidth{0pc}
\tablecaption{Spectral fitting results for simulated multiphase spectra}
\tablehead{
\colhead{}    &  \multicolumn{5}{c}{\Chandra} &   \colhead{}   &
\multicolumn{4}{c}{\XMM} \\
\cline{3-6} \cline{8-11} \\
\colhead{Sim.} & \colhead{Fit Range (keV)} & \colhead{$kT$ (keV)}   & \colhead{$Z$ (solar)}    & \colhead{$nH$ ($10^{21}$\,cm$^{-2}$)} & \colhead{$\chi^2$/dof} &
\colhead{}    & \colhead{$kT$ (keV)}   & \colhead{$Z$ (solar)}    & \colhead{$nH$ ($10^{21}$\,cm$^{-2}$)} & \colhead{$\chi^2$/dof}}
\startdata
A & 0.7--7.0 & $7.28\pm0.12$ & $0.37\pm0.02$ & $2.66\pm0.03$ & 439.1 / 427 && $7.36\pm0.06$ & $0.38\pm0.01$ & $2.65\pm0.02$ & 1294.6 / 1260 \\
A & 6.0--6.8 & $7.65\pm0.63$ & $0.39\pm0.08$ & \hbox{}~~$2.70$ (frozen) & 51.8 / 51 && $7.70\pm0.28$ & $0.39\pm0.04$ & \hbox{}~~$2.70$ (frozen) & 157.7 / 155 \\
B & 0.7--7.0 & $7.50\pm0.15$ & $0.38\pm0.03$ & $2.65\pm0.04$ & 443.4 / 427 && $7.67\pm0.09$ & $0.39\pm0.01$ & $2.61\pm0.02$ &  1324.4 / 1260 \\
B & 6.0--6.8 & $9.19\pm0.57$ & $0.54\pm0.10$ & \hbox{}~~$2.70$ (frozen) & 51.5 / 51 && $9.16\pm0.22$ & $0.53\pm0.04$ & \hbox{}~~$2.70$ (frozen) & 156.9 / 155 \\
C & 0.7--7.0 & $7.03\pm0.11$ & $0.49\pm0.03$ & $2.67\pm0.03$ & 455.7 / 427 && $7.07\pm0.06$ & $0.48\pm0.01$ & $2.67\pm0.02$ & 1331.4 / 1260  \\
C & 6.0--6.8 & $6.45\pm0.53$ & $0.41\pm0.08$ & \hbox{}~~$2.70$ (frozen) & 51.5 / 51 && $6.46\pm0.24$ & $0.39\pm0.04$ & \hbox{}~~$2.70$ (frozen) & 157.6 / 155 \\
\vspace{-0.6em}
\enddata
\label{tab:multiphase_sim}
\tablecomments{Input spectra comprise equally-weighted 5 \& 12 keV components with equal abundance of 0.3 solar (A), 0.2 \& 0.5 solar (B) and 
vice versa (C), respectively. Errors are the standard deviation of the 1000 Monte Carlo realisations in each case.}
\end{deluxetable*}
 

%% file: table4.tex
%



\begin{deluxetable*}{lcccccccc}
\tablecolumns{9}
\tablewidth{0pc}
\tablecaption{Data for the hot spots in the central ICM of A478}
\tablehead{
\colhead{Region} & \colhead{Distance$^a$ (kpc)} & \colhead{Cts$^b$} & \colhead{$kT$ (keV)} & 
 \colhead{$\chi^2$/dof} & \colhead{\rmsub{L}{X}$^c$ (10$^{42}$ erg/s)} & 
 \colhead{Vol$^d$ (10$^{68}$ cm$^3$)} & \colhead{$\Delta E$$^e$ ($10^{58}$ erg)} & 
 \colhead{\rmsub{t}{cool}$^f$ (yr)}
}
\startdata
A & 48--59 & 1054 & $12.1^{+2.6}_{-5.0}$ & 57.65/43 & 9.3 & 2.3 & $11\pm6$ & 6$\times10^8$ \\   
B & 34--42 & 727 & $9.6^{+2.5}_{-4.4}$ & 28.93/29 & 6.7 & 0.9 & $4.6\pm4$ & 4$\times10^8$ \\    
C & 38--51 & 1108 & $10.3^{+2.7}_{-3.6}$ & 45.08/45 & 10 & 2.6 & $8.7\pm6.5$ & 6$\times10^8$ \\ 
D & 45--55 & 639 & $10.3^{+3.2}_{-4.5}$ & 25.99/25 & 5.4 & 1.2 & $5.0\pm4.6$ & 5$\times10^8$ \\ 
\vspace{-0.6em}
\enddata
\label{tab:hotspot}
\tablecomments{$^a$Nearest/furthest distance to hotspot, measured from the cluster center; 
 $^b$net source counts in the 0.5--7.0 keV band; $^c$bolometric X-ray luminosity;
 $^d$estimated volume of region; $^e$estimated excess energy in hot spot; $^f$gas
 cooling time ($E$/\rmsub{L}{X}, using the \emph{total} rather than the excess hot spot energy). 
 All errors are 1$\sigma$.}
\end{deluxetable*}
 